\documentclass[iop]{emulateapj}
\usepackage{graphicx}
\usepackage{dcolumn}
\usepackage{bm}
\usepackage{float}
\usepackage{listings}
\usepackage{color}
\usepackage{textcomp}
\usepackage{times}
\usepackage{graphics}
\usepackage{amssymb}
\usepackage{latexsym}  
\usepackage{amsmath}
\usepackage{fancyhdr}
\usepackage[figuresright]{rotating}
\usepackage{hyperref}
\usepackage{array}
\usepackage{multirow}
\usepackage{afterpage}

\newcommand{\degree}{\ensuremath{^\circ}}
\newcommand{\pkssrc}{PKS~B0008-421 }
\newcommand{\pkssrcnospace}{PKS~B0008-421}
\shorttitle{Spectral Modeling of GPS Source PKS~B0008-421}
\shortauthors{\textsc{Callingham et al.}}
\submitted{Accepted to The Astrophysical Journal (ApJ) July 15, 2015}

\begin{document}
	
\title{Broadband Spectral Modeling of the Extreme Gigahertz-Peaked Spectrum Radio Source PKS~B0008-421}
	
\author{J.~R.~Callingham\altaffilmark{1,2,3}, B.~M.~Gaensler\altaffilmark{1,3,4}, R.~D.~Ekers\altaffilmark{2}, S.~J.~Tingay\altaffilmark{5,3}, R.~B.~Wayth\altaffilmark{5,3}, J.~Morgan\altaffilmark{5}, G.~Bernardi\altaffilmark{6,7}, M.~E.~Bell\altaffilmark{2,3}, R.~Bhat\altaffilmark{5,3}, J.~D.~Bowman\altaffilmark{8}, F.~Briggs\altaffilmark{9,3},  R.~J.~Cappallo\altaffilmark{10}, A.~A.~Deshpande\altaffilmark{11}, A.~Ewall-Wice\altaffilmark{12}, L.~Feng\altaffilmark{12}, L.~J.~Greenhill\altaffilmark{13}, B.~J.~Hazelton\altaffilmark{14}, L.~Hindson\altaffilmark{15}, N.~Hurley-Walker\altaffilmark{5}, D.~C.~Jacobs\altaffilmark{8}, M.~Johnston-Hollitt\altaffilmark{15}, D.~L.~Kaplan\altaffilmark{16}, N.~Kudrayavtseva\altaffilmark{5}, E.~Lenc\altaffilmark{1,3}, C.~J.~Lonsdale\altaffilmark{10},  B.~McKinley\altaffilmark{17,3}, S.~R.~McWhirter\altaffilmark{10}, D.~A.~Mitchell\altaffilmark{2,3}, M.~F.~Morales\altaffilmark{14}, E.~Morgan\altaffilmark{12}, D.~Oberoi\altaffilmark{18}, A.~R.~Offringa\altaffilmark{19,3}, S.~M.~Ord\altaffilmark{5,3}, B.~Pindor\altaffilmark{17,3}, T.~Prabu\altaffilmark{11}, P.~Procopio\altaffilmark{17,3},  J.~Riding\altaffilmark{17,3}, K.~S.~Srivani\altaffilmark{11}, R.~Subrahmanyan\altaffilmark{11,3}, N.~Udaya~Shankar\altaffilmark{11},  R.~L.~Webster\altaffilmark{17,3}, A.~Williams\altaffilmark{5}, C.~L.~Williams\altaffilmark{12}}

\affil{$^1$Sydney Institute for Astronomy (SIfA), School of Physics, The University of Sydney, NSW 2006, Australia}
\affil{$^2$CSIRO Astronomy and Space Science (CASS), Marsfield, NSW 2122, Australia}
\affil{$^3$ARC Centre of Excellence for All-Sky Astrophysics (CAASTRO)}
\affil{$^4$Dunlap Institute for Astronomy \& Astrophysics, University of Toronto, Toronto, ON, M5S 3H4, Canada}
\affil{$^5$International Centre for Radio Astronomy Research (ICRAR), Curtin University, Bentley, WA 6102, Australia}
\affil{$^6$Square Kilometre Array South Africa (SKA SA), Pinelands, 7405, South Africa}
\affil{$^7$Department of Physics and Electronics, Rhodes University,Grahamstown, 6140, South Africa}
\affil{$^8$School of Earth and Space Exploration, Arizona State University, Tempe, AZ 85287, USA}
\affil{$^9$Research School of Astronomy and Astrophysics, Australian National University, Canberra, ACT 2611, Australia}
\affil{$^{10}$MIT Haystack Observatory, Westford, MA 01886, USA}
\affil{$^{11}$Raman Research Institute, Bangalore 560080, India}
\affil{$^{12}$Kavli Institute for Astrophysics and Space Research, Massachusetts Institute of Technology, Cambridge, MA 02139, USA}
\affil{$^{13}$Harvard-Smithsonian Center for Astrophysics, Cambridge, MA 02138, USA}
\affil{$^{14}$Department of Physics, University of Washington, Seattle, WA 98195, USA}
\affil{$^{15}$School of Chemical \& Physical Sciences, Victoria University of Wellington, Wellington 6140, New Zealand}
\affil{$^{16}$Department of Physics, University of Wisconsin--Milwaukee, Milwaukee, WI 53201, USA}
\affil{$^{17}$School of Physics, The University of Melbourne, Parkville, VIC 3010, Australia}
\affil{$^{18}$National Centre for Radio Astrophysics, Tata Institute for Fundamental Research, Pune 411007, India}
\affil{$^{19}$Netherlands Institute for Radio Astronomy (ASTRON), Dwingeloo, The Netherlands}

\email{j.callingham@physics.usyd.edu.au}
	
\begin{abstract}
		
We present broadband observations and spectral modeling of \pkssrcnospace, and identify it as an extreme gigahertz-peaked spectrum (GPS) source. \pkssrc is characterized by the steepest known spectral slope below the turnover, close to the theoretical limit of synchrotron self-absorption, and the smallest known spectral width of any GPS source. Spectral coverage of the source spans from 0.118 to 22\,GHz, which includes data from the Murchison Widefield Array and the wide bandpass receivers on the Australia Telescope Compact Array. We have implemented a Bayesian inference model fitting routine to fit the data with internal free-free absorption, single and double component free-free absorption in an external homogeneous medium, free-free absorption in an external inhomogeneous medium, or single and double component synchrotron self-absorption models, all with and without a high-frequency exponential break. We find that without the inclusion of a high-frequency break these models can not accurately fit the data, with significant deviations above and below the peak in the radio spectrum. The addition of a high-frequency break provides acceptable spectral fits for the inhomogeneous free-free absorption and double-component synchrotron self-absorption models, with the inhomogeneous free-free absorption model statistically favored. The requirement of a high-frequency spectral break implies that the source has ceased injecting fresh particles. Additional support for the inhomogeneous free-free absorption model as being responsible for the turnover in the spectrum is given by the consistency between the physical parameters derived from the model fit and the implications of the exponential spectral break, such as the necessity of the source being surrounded by a dense ambient medium to maintain the peak frequency near the gigahertz region. This implies that \pkssrc should display an internal H\,\textsc{i} column density greater than $10^{20}$\,cm$^{-2}$. The discovery of \pkssrc suggests that the next generation of low radio frequency surveys could reveal a large population of GPS sources that have ceased activity, and that a portion of the ultra-steep spectrum source population could be composed of these GPS sources in a relic phase. 

\end{abstract}
	
\keywords{galaxies: active --- galaxies: individual (\pkssrcnospace) --- radiation mechanisms: general --- radio continuum: general --- radio sources: spectra}	
	
\section{Introduction}

Gigahertz-peaked spectrum (GPS) radio sources are a class of active galactic nuclei (AGN) that have played a pivotal role in shaping our understanding of the environmental properties and evolutionary paths of radio galaxies. These powerful radio sources are defined by a concave spectrum that peaks at gigahertz frequencies, steep spectral slopes on either side of the turnover, small linear sizes ($\sim$\,$0.1 - 1$\,kpc), minimal variability over timescales varying from hours to decades, and low radio polarization fractions \citep{Odea1991}. Two closely related groups of radio sources are high frequency peakers (HFPs) and compact steep spectrum (CSS) sources, which display many of the same physical properties as the GPS population but with different peak frequencies and linear sizes. HFPs peak above a gigahertz and have smaller linear sizes than GPS sources ($\lesssim$\,0.1\,kpc) \citep{2000A&A...363..887D,2005A&A...432...31T,2010MNRAS.408.1187H}, while CSS sources peak below gigahertz frequencies and have larger linear sizes than GPS sources ($\sim$\,$1 - 10$\,kpc) \citep{1990A&A...231..333F,Fanti1995}. The hosts of HFP, GPS and CSS sources can be associated with quasars, radio galaxies, or Seyfert galaxies \citep{1994ApJS...91..491G,deVries1997,2003PASA...20..118S}. A comprehensive review of GPS and CSS sources has been presented by \citet{Odea1998}, with additional and more recent research summaries provided by \citet{Stawarz2008} and \citet{Marr2014}.

Very long baseline interferometry (VLBI) imaging of HFP, GPS and CSS sources has allowed these radio sources to be further split into two distinct classes based on their morphology. The first morphological class is defined by a core-jet structure, and such sources are generally associated with high redshift quasars \citep{Stanghellini1997,2006A&A...450..959O}. The second morphological class of GPS and CSS sources is characterized by dominant small scale structures that are reminiscent of the large scale radio lobes of powerful radio galaxies, with inverted or flat spectrum cores surrounded by two steep-spectra lobes. GPS and CSS sources that display these characteristics are referred to as compact symmetric objects (CSOs), and are generally associated with low redshift radio galaxies \citep{Wilkinson1994,2009A&A...498..641D}. CSOs have been the focus of extensive study because such morphologies suggest that these sources could be the young precursors to large radio galaxies \citep{Marr2014}.

This `youth' hypothesis \citep{Phillips1982,Fanti1995,2000MNRAS.319..445S} has observational support from hotspot measurements \citep{Owsianik1998,Polatidis2003,Gugliucci2005} and from high-frequency spectral break modeling \citep{Murgia1999,Orienti2010}. However, such an interpretation is still contentious, as statistical studies of the luminosity functions have demonstrated an over abundance of CSOs relative to the number of large radio galaxies \citep{Odea1997,Readhead1996,An2012}. An alternative hypothesis is the `frustration' model, which implies that these sources are not young but are confined to small spatial scales due to unusually high nuclear plasma density \citep{vanBreugel1984}. It is possible that both these scenarios may apply to GPS sources, as young sources with constant AGN activity could break through a dense medium given sufficient time, while sources with intermittent AGN activity may stagnate \citep{An2012}.

One of the reasons that there has not been a resolution between these two competing hypotheses is because the absorption mechanism responsible for the spectral turnover still remains an open area of debate. Synchrotron self-absorption (SSA) by the relativistic electrons internal to the emitting source, and free-free absorption (FFA) via an external ionized screen of dense plasma, are two commonly proposed models for the turnover \citep{Kellermann1966}. Both these absorption models replicate some physical features of GPS sources but fail to explain others. For example, the observed correlation between the turnover frequencies and linear sizes of GPS sources is well justified within the SSA framework, while FFA via a homogeneous medium can not replicate such a relationship \citep{Odea1998}. However, FFA models that invoke an external inhomogeneous medium where electron density decreases with distance from the radio jet \citep[e.g.][]{Bicknell1997,Kuncic1998}, or dense ionized interstellar clouds that co-exist with the relativistic electrons \citep[e.g.][]{Begelman1999,Stawarz2008,Maciel2014}, are able to recreate this correlation. While it has been shown that it is likely that this `frustration' sceanrio is inconsistent with the properties of many GPS sources \citep{2000MNRAS.319..445S,2003PASA...20...38S,2009AN....330..214D}, there is mounting evidence from many observational studies of individual GPS sources, that have morphologies consistent with CSOs, demonstrating FFA is responsible for the inverted spectra \citep{Peck1999,Kameno2000,Marr2001,Orienti2008,2008ApJ...684..153T,Marr2014,Tingay2015}.

It is possible to differentiate between FFA and SSA models through comprehensive statistical fitting of GPS spectra, provided the spectra are well sampled below the turnover. For example, \citet{Tingay2003} performed a detailed investigation of different absorption scenarios via model fitting to the radio spectrum of the GPS source PKS B1718-649. They concluded that SSA was the most likely contributor to the inverted spectrum, but emphasized that the modified inhomogeneous FFA model of \citet{Bicknell1997} could not be excluded. The reason the degeneracy between SSA and FFA models has remained unresolved is because these past investigations lacked comprehensive spectral coverage below the turnover, where the distinction between the different absorption models becomes pronounced. With new low radio frequency telescopes such as the Murchison Widefield Array \citep[MWA;][]{Lonsdale2009,Tingay2013} and the LOw-Frequency ARray \citep[LOFAR;][]{vanHaarlem2013} becoming operational, astronomers now have unprecedented frequency coverage below the turnover, allowing for a more comprehensive spectral comparison of the different absorption models. Additionally, the new wide bandpass receivers installed on existing radio telescopes, such as the Compact Array Broadband Backend \citep[CABB;][]{Wilson2011} on the Australia Telescope Compact Array (ATCA), provide new broadband spectral coverage at gigahertz frequencies. Such broadband spectral coverage is also vital to understanding the nature of high-frequency spectral breaks in GPS sources, which can heavily influence spectral fits below the turnover \citep{Tingay2015}.

In light of these technological advancements, we present MWA and CABB observations of GPS source \pkssrc (RA = 00:10:52.5, Dec. = --\,41:53:10.6 (J2000); $z$ = 1.12; \citealt{Labiano2007}). \pkssrc was selected for this study because it has the steepest spectral slope below the turnover for sources observed during the MWA Commissioning Survey \citep[MWACS;][]{Hurley-Walker2014}. \pkssrc was identified as a candidate GPS source by \citet{Odea1991}. While several studies have demonstrated properties of \pkssrc that are consistent with a GPS source classification \citep{Labiano2007,Jacobs2011}, we present the first flux density measurements below the turnover that confirm this characterization. VLBI measurements of \pkssrc reveal a morphology that is consistent with a CSO \citep{King1994,Jauncey2003}. Furthermore, \pkssrc has extensive temporal information and wide spectral coverage from 0.118 to 22\,GHz, since it has acted as a primary or secondary calibrator for most southern hemisphere radio telescopes, allowing us to overcome the epochal and spectral limitations of previous studies of GPS sources. 

The purpose of this paper is to present unprecedented spectral coverage of this extreme GPS source, and to demonstrate that advances in wide bandpass receivers and low radio frequency coverage now allow us to place stringent observational constraints on the absorption models and physical environments of GPS sources. The relevant data reduction procedures performed for the MWA, CABB and archival observations are discussed in \S\,\ref{datared}. In \S\,\ref{fittingroutine} the Bayesian fitting routine that we implemented to assess the different absorption model fits to \pkssrc is outlined. Relevant features of the absorption models, and their respective fitting statistics, are presented in \S\,\ref{results}. In \S\,\ref{discussion} the impact the absorption models fits have on our understanding of the environment of \pkssrcnospace, and the absorption models used to describe GPS sources as a whole, are discussed. 

\section{Observations, Data Reduction and Archival Data}\label{datared}

\pkssrc was regularly observed between 1969 and 2014 by a variety of radio telescopes. Most of the observations have been performed by the Parkes 64\,m radio telescope, the ATCA, the Molonglo Observatory Synthesis Telescope (MOST) and the Giant Metrewave Radio Telescope (GMRT). Combining all these observations with the MWA and ATCA observations provides a spectral coverage of \pkssrc from 0.118 to 22\,GHz. These observations are summarized in Table \ref{fluxvals} and plotted in Figure \ref{nomodelsed}. \pkssrc was spatially unresolved in all these observations.  
\begin{table*}
	\small
	\caption{\label{fluxvals} A summary of the observations of \pkssrc used in the spectral modeling. Note that the reported \citet{Wills1975} fluxes have been corrected to the \citet{Baars1977} flux density scale. The epochs of the observations are presented as accurately as possible, but when only a broad timeframe is known, it is presented in the format YYYY-YYYY.}

	\begin{center}
		\begin{tabular}{ccclll}
		\hline
		\hline
$\nu$ & $S_{\nu}$ & $\sigma$ & Epoch & Telescope & Reference (Survey)\\
(GHz) & (Jy) & (Jy) & & & \\
		\hline			
		0.118 & 0.7 & 0.2 & 2012 Dec 01     & MWA      & \citet{Hurley-Walker2014} (MWACS)\\
		0.150 & 1.2 & 0.2 & 2012 Dec 01     & MWA      & \citet{Hurley-Walker2014} (MWACS)\\
		0.180 & 1.9 & 0.1 & 2012 Dec 01     & MWA      & \citet{Hurley-Walker2014} (MWACS)\\
		0.235 & 3.6 & 0.5 & 2013 Jul 04  & GMRT     & GMRT Cal\\
		0.408 & 6.6 & 0.2 & 1969-1974  & MOST     & \citet{Large1981} (MRC)\\
		0.468 & 7.0 & 0.7 & 1965-1969  & Parkes Interferometer & \citet{Wills1975}\\
		0.580 & 6.6 & 0.4 & 1965-1969  & Parkes   & \citet{Wills1975}\\
		0.610 & 6.7 & 1.1 & 2013 Jul 04 & GMRT     & GMRT Cal\\
		0.635 & 7.3 & 0.2 & 1965-1969  & Parkes   & \citet{Wills1975}\\
		0.843 & 6.5 & 0.3 & 1987 Nov 07 & MOST     & \citet{CampellWilson1994}\\
		0.843 & 6.4 & 0.2 & 1998 Apr 05  & MOST     & \citet{Mauch2003} (SUMSS)\\
		0.960 & 6.1 & 0.2 & 1965-1969  & Parkes   & \citet{Wills1975}\\
		1.357 & 4.6 & 0.1 & 2012 Jun 15 & ATCA     &  This work\\
		1.384 & 4.4 & 0.4 & 2008 Jun 19 & ATCA     & \citet{Randall2011}\\
		1.410 & 4.70 & 0.07 & 1965-1969  & Parkes   & \citet{Wills1975}\\
		1.687 & 3.83 & 0.09 & 2012 Jun 15 & ATCA     & This work\\
		1.904 & 3.43 & 0.08 & 2012 Jun 15 & ATCA     & This work\\
		2.106 & 3.14 & 0.08 & 2012 Jun 15 & ATCA     & This work\\
		2.307 & 2.87 & 0.07 & 2012 Jun 15 & ATCA     & This work\\
		2.496 & 2.7 & 0.3 & 2008 Jun 19   & ATCA      & \citet{Randall2011}\\
		2.513 & 2.65 & 0.06 & 2012 Jun 15 & ATCA     & This work\\
		2.700 & 2.49 & 0.03 & 1965-1969  & Parkes   & \citet{Wills1975}\\
		2.718 & 2.45 & 0.06 & 2012 Jun 15 & ATCA     & This work\\
		2.921 & 2.28 & 0.05 & 2012 Jun 15 & ATCA     & This work\\
		4.680 & 1.37 & 0.03 & 2010 Jun 10 & ATCA     & This work\\
		4.800 & 1.35 & 0.07 & 2004 Nov 11  & ATCA     & \citet{Murphy2010} (AT20G) \\
		4.800 & 1.25 & 0.02 & 1993 Sep 24  & ATCA     & \citet{McConnell2012} (ATPMN) \\
		4.800 & 1.3 & 0.1 & 2008 Jun 19   & ATCA     & \citet{Randall2011}\\
		4.850 & 1.39 & 0.07 & 1990 Jun 01  & Parkes   & \citet{Wright1994} (PMN)\\
		4.926 & 1.28 & 0.03 & 2010 Jun 10 & ATCA     & This work \\
		5.000 & 1.1 & 0.2 & 1965-1969  & Parkes   & \citet{Wills1975}\\
		5.009 & 1.35 & 0.08 & 1965-1969  & Parkes   & \citet{Wills1975}\\
		5.145 & 1.22 & 0.03 & 2010 Jun 10 & ATCA     & This work\\
		5.380 & 1.75 & 0.03 & 2010 Jun 10 & ATCA     & This work\\
		5.615 & 1.09 & 0.03 & 2010 Jun 10 & ATCA     & This work\\
		5.822 & 1.04 & 0.03 & 2010 Jun 10 & ATCA     & This work\\
		6.062 & 0.98 & 0.02 & 2010 Jun 10 & ATCA     & This work\\
		6.264 & 0.94 & 0.02 & 2010 Jun 10 & ATCA     & This work\\
		8.215 & 0.65 & 0.02 & 2010 Jun 10 & ATCA     & This work\\
		8.478 & 0.62 & 0.01 & 2010 Jun 10 & ATCA     & This work\\
		8.640 & 0.57 & 0.06 & 2008 Jun 19  & ATCA     & \citet{Randall2011}\\
		8.640 & 0.55 & 0.03 & 1993 Sep 24  & ATCA     & \citet{McConnell2012} (ATPMN)\\
		8.640 & 0.61 & 0.03 & 2004 Nov 22  & ATCA     & \citet{Murphy2010} (AT20G)\\
		8.690 & 0.60 & 0.02 & 2010 Jun 10 & ATCA     & This work\\
		8.936 & 0.58 & 0.02 & 2010 Jun 10 & ATCA     & This work\\
		9.156 & 0.56 & 0.01 & 2010 Jun 10 & ATCA     & This work\\
		9.360 & 0.55 & 0.01 & 2010 Jun 10 & ATCA     & This work\\
		9.575 & 0.54 & 0.01 & 2010 Jun 10 & ATCA     & This work\\
		9.776 & 0.52 & 0.01 & 2010 Jun 10 & ATCA     & This work\\
		18.500 & 0.18 & 0.01 & 2002 Mar 21 & ATCA    & \citet{Ricci2004}\\
		19.904 & 0.152 & 0.008 & 2004 Oct 22 & ATCA    & \citet{Murphy2010} (AT20G)\\
		22.000 & 0.13 & 0.01 & 2001 Jan 02  & ATCA    & \citet{Ricci2006}\\
                                                      
		\hline\end{tabular}                           
\end{center}                                                                               
\end{table*}

\subsection{ATCA Observations and Data Reduction}

As part of project C007 (PI Stevens), \pkssrc was observed by the ATCA in the 6C array configuration on 2010 June 9-10. The observation was conducted using the CABB backend system, which gives two instantaneous 2\,GHz bandwidth for both linear polarizations, at central frequencies 5.5\,GHz and 9.0\,GHz for a total integration time of 32 minutes. Each observation used 1\,MHz channels and a 10 second correlator integration time. \pkssrc was also observed for project C2697 (PI Ryder) with the ATCA in the 6D configuration on 2012 June 14-15. This observation was conducted at a central frequency of 2.1\,GHz and for a total integration time of 121 minutes, using the same bandwidth, channel size and correlator integration time as the 2010 observation. PKS~B1934-638 was used for flux density, bandpass and phase calibration for both observations.

The data for both observations were reduced using the \textsc{miriad} software package \citep{Sault1995}. The known regions of radio frequency interference (RFI) and lower sensitivity in the CABB system were initially flagged. RFI was excised for PKS~B1934-638 using the automatic flagging option in $\tt{pgflag}$ and manually with $\tt{blflag}$. The bandpass, gain and leakage solutions were estimated using PKS~B1934-638. The gain calibration was performed over four 512\,MHz wide frequency bins and ten second time intervals. The calibration solutions were transferred to \pkssrc and the flux density scale was set to that defined by PKS~B1934-638.

The flux density of \pkssrc was measured in every spectral channel of the visibilities using the task $\tt{uvfmeas}$. The uncertainty of each flux density measurement is dominated by thermal noise and gain calibration errors.  The thermal noise was estimated by dividing the visibility space into a large number of independent cells of a size given by a window function, and by using the physical antenna size, bandwidth, total system temperature, time averaged number of baselines in the cell, and total observation time \citep[Eqn. 11;][]{Morales2005}. The thermal noise is Gaussian and independent between the channels \citep{Bowman2006,Wilson2011}. The gain calibration errors were estimated by examining the variation of flux density of the source and calibrator over different baselines and by comparison of the root-mean-square residuals between the measured visibilities and the model visibilities of a point source. The gain calibration errors are independent since the CABB continuum channels have a `square' response \citep[Fig. 12;][]{Wilson2011} but may not be normally distributed since they are antenna dependent. However, the magnitude of the deviation from a Gaussian distribution is dependent on the uniformity of the gain response of the antennas in the array \citep{Boonstra2003}. Since the gain response of the antennas of the ATCA are within $\sim$\,1\% of each other \citep{Wilson2011}, it can be assumed that the uncertainties are approximately normally distributed. Additionally, the effect of confusing sources on the spectrum of \pkssrc is negligible, with the next brightest source within the full width at half maximum of the primary beam of flux density between $\sim$\,0.4 to 0.6\,mJy for all observing frequencies. For the same frequency range, \pkssrc varies between $\sim$\,4 to 0.5\,Jy.

The flux density measurements and their associated uncertainties were grouped into bins of approximately 200\,MHz, with variations in bin size occurring since sections of the spectrum had been flagged due to RFI. Images were also produced at 1.8, 5.5 and 9.0\,GHz to ensure that the visibilities of \pkssrc were not contaminated by confusing sources and to confirm that \pkssrc was spatially unresolved at all frequencies. 

\subsection{MWA Observations and Data Reduction}

The MWA observed \pkssrc during MWACS, a survey which covers approximately 21\,h $<$ RA $<$ 8\,h, $-$\,55$\degree$ $<$ Dec. $<$ $-$\,10$\degree$ over three frequency bands centered at 0.199, 0.150 and 0.180\,GHz \citep{Hurley-Walker2014}. MWACS was conducted between September and December 2012 by taking drift scans at declinations approximately $-$\,27$\degree$ and $-$\,47$\degree$, using two different 32-tile subarrays of the full 128-tile array (where a tile consists of 16 dual-polarization dipole antennas). The survey has a 3 arcminute resolution and noise level of $\sim$\,40\,mJy\,beam$^{-1}$ at 0.150\,GHz.  

Given that the MWA flux density measurements are vital in constraining the spectral slope below the turnover of \pkssrcnospace, it is important to summarize the absolute flux density calibration that was undertaken for MWACS. The absolute flux density scale was empirically evaluated by bootstrapping from a source that had been observed by northern and southern hemisphere radio telescopes. The source 3C32 was selected since it has wide spectral coverage that is well fit by a simple power-law, shows no evidence of variability, and is unresolved in MWACS and in the source catalogs used to fill in its spectral energy distribution. The flux densities of sources in the Dec.~$-$\,27$\degree$ maps were corrected by the least-squares fit to the catalog flux densities of 3C32. The Dec.~$-$\,47$\degree$ mosaic was corrected to the flux density scale of the Dec.~$-$\,27$\degree$ mosaic by using $\approx$\,600 unresolved sources in the overlapping region of the two mosaics (see \S\,3.4 of Hurley-Walker et al. 2014 for more details). \pkssrc was unresolved in all images that contributed to MWACS. Note that we have used all three flux densities measured by MWACS, while the published catalog of \citet{Hurley-Walker2014} only reports a flux density at 0.180\,GHz based on a spectral fit to the flux density at the three frequencies. Using all three flux density measurements is justified due to the brightness of \pkssrc relative to other sources in the catalog. Confusion also has a negligible impact on the flux density measurements at these frequencies since \pkssrc is at least four times brighter than any other source in the synthesized primary beam.

\subsection{Archival Data}

\subsubsection{ATCA}

\pkssrc has been observed by the ATCA as part of both survey and targeted programs. For targeted observations, \citet{Randall2011} observed \pkssrc as part of a campaign to construct an unbiased southern hemisphere catalog of CSS and GPS sources. They conducted observations at 1.4, 2.3, 4.8 and 8.6\,GHz in the 1.5B array using the now superseded 128\,MHz array backend. The source was not included as part of their final catalog since the observed frequencies were not low enough to constrain the turnover. \pkssrc was also observed by \citet{Ricci2004} as part of a program focusing on the polarization properties of bright radio sources at 18.5\,GHz. Since the 15\,mm receivers were in the commissioning phase at the time of observation, only three antennas in a compact configuration were used with 128\,MHz bandwidth. Additionally, \citet{Ricci2004} had poor atmospheric phase stability during the observations so they used non-imaging model-fitting techniques to calibrate and derive source flux densities in all the Stokes parameters after accounting for confusion. The sources of \citet{Ricci2004} were followed up by \citet{Ricci2006} at 22\,GHz.

During the blind Australia Telescope 20\,GHz Survey \citep[AT20G;][]{Murphy2010}, \pkssrc was observed with the ATCA using two 128\,MHz bands centered on 18.752 and 21.056\,GHz, which were then combined into a single band centered on 19.9\,GHz. The source was followed up at 4.8 and 8.6\,GHz. Upper limits were placed on the polarized flux and fractional polarization at these frequencies. All observations were conducted in a hybrid array. \pkssrc was also observed as part the Australia Telescope-Parkes-MIT-NRAO survey \citep[ATPMN;][]{McConnell2012}, which was an ATCA survey of Parkes-MIT-NRAO \citep[PMN;][]{Wright1994} sources at 4.8 and 8.64\,GHz, with the array in 6C and using a 128\,MHz bandpass. \pkssrc was spatially unresolved for all observations conducted by the ATCA. Furthermore, \pkssrc remained unresolved when the AT20G data were reprocessed utilizing the study of 6\,km baselines \citep{Chhetri2013}. All ATCA flux densities are listed in Table \ref{fluxvals}.

\subsubsection{Parkes 64\,m Radio Telescope}
 
The Parkes 64\,m radio telescope observed \pkssrc between 1965 and 1969 at 0.580, 0.635, 0.960, 1.410, 2.700, 5.000 and 5.009\,GHz \citep[and references therein]{Wills1975}. At 0.580, 0.635 and 0.960\,GHz uncertainty due to confusion can be large since the Parkes 64\,m radio telescope is a single dish, so scans were made over a large range of right ascension and declination to correct for effects of any strong confusing sources on the base level on either side of the scan. \pkssrc is located in a field that is isolated from sources brighter than 1\,Jy at 0.635\,GHz. \pkssrc was also observed by the Parkes interferometer, which consisted of the 64\,m radio telescope and a 18\,m radio telescope, at 0.468\,GHz. This had the advantage of allowing the effects of confusing sources on source structure to be separated and removed. These observations are discussed at greater length by \citet{Wills1975}. Note that the flux densities have been adjusted from the flux density scale of \citet{Wills1973} to the \citet{Baars1977} flux density scale via appropriate frequency dependent multiplicative factors. Additionally, \pkssrc was observed as part of the PMN survey at 4.85\,GHz \citep{Wright1994}. The Parkes 64\,m radio telescope flux density measurements are also listed in Table \ref{fluxvals}.

\subsubsection{Molonglo Observatory Synthesis Telescope}

\pkssrc was observed by the MOST during the Sydney University Molonglo Sky Survey \citep[SUMSS;][]{Bock1999,Mauch2003} at 0.843\,GHz and the Molonglo Reference Catalogue  \citep[MRC;][]{Large1981} at 0.408\,GHz. Additionally, \pkssrc was frequently observed by the MOST at 0.843\,GHz as part of a calibrator observation campaign from 1987 to 1996 \citep{CampellWilson1994,Gaensler2000}. The flux density measurements are presented in Table \ref{fluxvals}. See the discussion on variability in \S\,\ref{variability} for further details about these observations. 

\subsubsection{Giant Metrewave Radio Telescope}

The GMRT observed \pkssrc at 0.235 and 0.610\,GHz as part of its observing program to identify suitable low radio frequency calibrators\footnote{Details on the data reduction process can be found at \url{http://gmrt.ncra.tifr.res.in/gmrt_hpage/Users/Help/CAL/Cal-List.html}}. \pkssrc was unresolved at both frequencies. Uncertainties on the flux density measurements were calculated from the noise in the images and the errors in the gain solution of PKS~B1934-638. The flux densities are presented in Table \ref{fluxvals}.

\subsubsection{Very Long Baseline Interferometry Observations}\label{VLBIsec}

VLBI observations of GPS sources are integral to understanding their nature because it is only with this high level of resolution that the source structure can be characterized. \pkssrc was part of a VLBI monitoring campaign of southern hemisphere GPS sources at 2.3 and 8.4\,GHz from 1990 to 1993 using a combination of antennas at seven different locations in Australia and one in South Africa \citep{King1994,Jauncey2003}. Based on model fitting to the visibilities and closure phases, \pkssrc was well described by a model comprising two aligned linear components \citep{King1994}. At 2.3\,GHz, the separation of these two components was observed to be $\sim$\,120 milliarcseconds, which at $z = 1.12$ corresponds to a physical separation of $\sim$\,1000 pc; calculated assuming a spatially flat $\lambda$CDM cosmology with matter density $\Omega_{\mathrm{M}} = 0.286$, vacuum energy density $\Omega_{\mathrm{\lambda}} = 0.714$, and Hubble constant $H_{0} = 69.6$\,km\,s$^{-1}$ Mpc$^{-1}$ \citep{2006PASP..118.1711W}. These structural properties are also evident in VLBA measurements of the source at 8.6\,GHz \citep{Fey2004}, although detailed model fitting was not conducted for these observations. This classical double structure and small separation makes \pkssrc consistent with the morphological CSO classification of GPS sources. 

\begin{figure*}
\begin{center}
\includegraphics[scale=0.45]{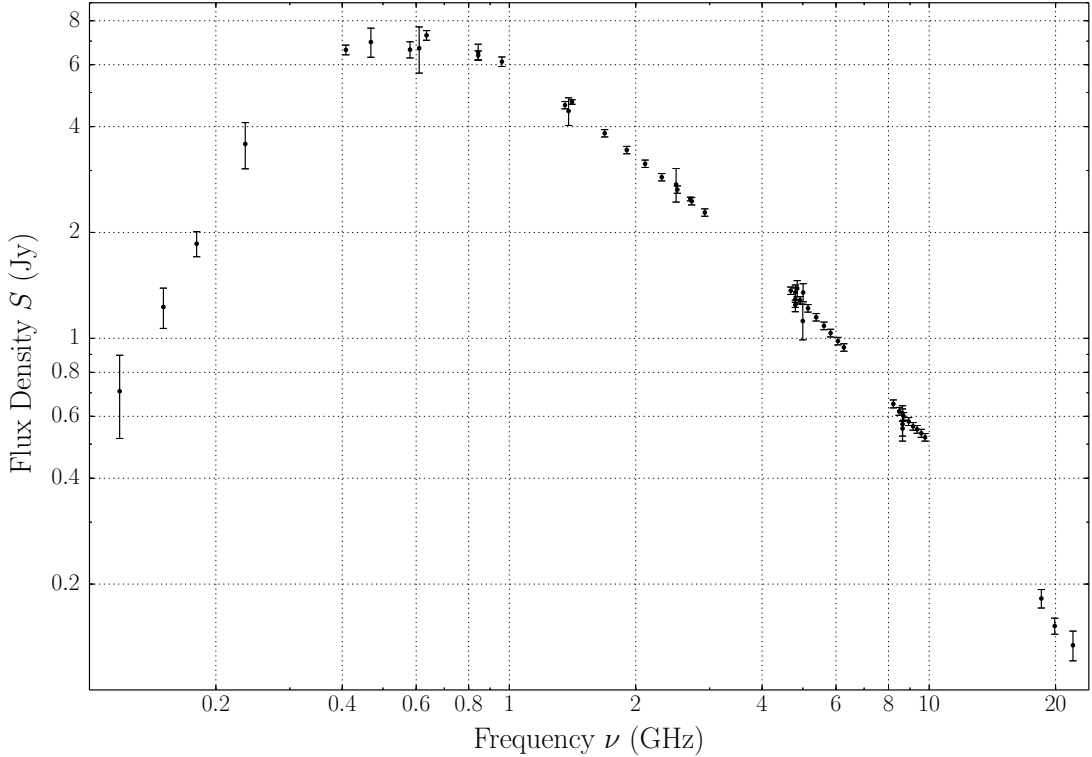}
\caption{Spectral energy distribution for \pkssrcnospace, as described by the data points in Table \ref{fluxvals}. All points plotted are in the observed frame of \pkssrcnospace.}
\label{nomodelsed}
\end{center}
\end{figure*}

\section{Bayesian Inference Model Fitting Routine}\label{fittingroutine}

We have implemented a Bayesian inference model fitting routine to assess the extent to which different absorption models can describe the spectrum of \pkssrcnospace, and to calculate the respective model parameters and associated uncertainties. Suppose that the observed flux densities at various frequencies are represented by the data vector $\bf{D}$ and that we want to estimate the values of the parameters $\boldsymbol{\theta}$ in some underlying model of the data $M$. One possible way to approach this problem is to utilize Bayes' theorem,

\begin{equation}\label{Bayeseqn}
	\mathrm{Pr}(\boldsymbol{\theta}\,|\,\boldsymbol{D},M) = \dfrac{\mathrm{Pr}(\boldsymbol{D}\,|\,\boldsymbol{\theta},M)\mathrm{Pr}(\boldsymbol{\theta}\,|\,M)}{\mathrm{Pr}(\boldsymbol{D}\,|\,M)},
\end{equation}

\noindent where $\mathrm{Pr}(\boldsymbol{\theta}\,|\,\boldsymbol{D},M)$ is the posterior distribution of the model parameters, representing the updated belief of the model parameters given the data, $\mathrm{Pr}(\boldsymbol{D}\,|\,\boldsymbol{\theta},M) \equiv \mathcal{L}(\boldsymbol{\theta})$ is the likelihood function, which is the probability of observing the data provided given some model parameters, $\mathrm{Pr}(\boldsymbol{\theta}\,|\,M) \equiv \Pi(\boldsymbol{\theta})$ is the prior information associated with the model, and the Bayesian evidence $\mathrm{Pr}(\boldsymbol{D}\,|\,M) \equiv Z$ is a measure of how well the model predicts the data observed.

The flux density measurements of \pkssrc in Table \ref{fluxvals} are independent since the measurements are conducted by different instruments or the receivers had a response that ensures independence between channels. We make the justified assumption that the uncertainty $\sigma_{n}$ on each measurement is normally distributed. 

Since the flux density measurements are independent and the uncertainty on each measurement is normally distributed, the joint log likelihood function we used for this analysis was: 

\begin{equation}\label{likelihoodeqn}
	\ln \mathcal{L}(\boldsymbol{\theta}) = -\dfrac{1}{2}\sum\limits_{n}\left[\dfrac{(D_{n} - S_{\nu}(\nu_{n}))^{2}}{\sigma_{n}^{2}} + \ln(2\pi\sigma_{n}^{2})\right],
\end{equation}

\noindent where $D_{n}$ is the flux density observed at frequency $\nu_{n}$ and $S_{\nu}(\nu_{n})$ is the expected flux density at $\nu_{n}$ from the model given parameters $\boldsymbol{\theta}$. This likelihood function was convolved with uniform priors for each model parameter with sensible ranges placed by known physical constraints. For example, the peak frequency or normalization of the flux density can not be negative. 

The most efficient way to calculate the posterior probability density functions for the parameters of different models is via Markov chain Monte Carlo (MCMC) methods, which are designed to discretely sample posterior probability density functions of model parameters such that the likelihood function is maximized. We implemented an affine-invariant ensemble sampler \citep{Goodman2010}, via the Python package $\tt{emcee}$ \citep{ForemanMackey2013}, which means the algorithm utilizes an ensemble of ``walkers'' and is relatively insensitive to covariance between parameters since it performs equally well under all linear transformations. We utilized the simplex algorithm to direct the walkers to the maximum of the likelihood function \citep{Nelder1965}. 

A Bayesian inference approach to modeling radio spectra has several advantages over traditional probabilistic data analysis procedures, such as least-squares and maximum-likelihood methods. This includes the ability to marginalize over nuisance parameters, generate complete probability distributions of model parameters and to incorporate prior knowledge so known physical constraints can be naturally placed on the system. Most importantly, since we are fitting non-linear models, Bayesian inference facilitates a more objective model comparison than traditional reduced $\chi^{2}$-tests via comparison of evidence values, which appropriately penalize any additional degrees of freedom by integrating over all parameter space in each model \citep{Jeffreys1961,Andrae2010}. 

Bayesian evidence is required to normalize the posterior over the prior volume so that:

\begin{equation}\label{evidence}
 Z = \int\int \cdots \int \mathcal{L}(\boldsymbol{\theta})\Pi(\boldsymbol{\theta})\mathrm{d}\boldsymbol{\theta},
\end{equation}

\noindent where the dimensionality of the integration is set by the number of parameters in the model. The evidence can be thought of as the numerical statement of Occam's razor, since it represents the average of the likelihood over the prior volume. This means that models with high likelihood values throughout parameter space are favored over models with low likelihood regions. Unless a complex model provides a significantly better fit to the data than a model with a smaller parameter space, the evidence value will be larger for the model with a smaller number of parameters. We calculated the evidence for the different models using the algorithm \textsc{MultiNest} \citep{Feroz2013}, which is an implementation of nested sampling. The algorithm is initialized by uniformly sampling the prior space and then contracting the distribution of samples around high likelihood regions by discarding the sample with the least likelihood. A random sample is duplicated to keep the number of samples constant. This process is repeated until the replacement of samples is optimized after finding the region of maximum likelihood. The strength of nested sampling is that it calculates the mean posterior probability and the evidence. We ensured that the parameter estimates from \textsc{MultiNest} were within uncertainty of those outputted by $\tt{emcee}$, such that the calculated evidence was indicative of the best model fit.

Assuming \emph{a priori} that two models are equally likely to describe the data, model selection can be performed solely based on the ratios of the respective evidences. In this paper we perform this in log space such that $\Delta\ln(Z) = \ln(Z_{2}) - \ln(Z_{1})$. Using an updated version of the Jefferys scale \citep*[e.g.][]{Kass1995,Scaife2012}, $\Delta\ln(Z) \geq 3$ is strong evidence that $M_{2}$ is favored over $M_{1}$, $1 < \Delta\ln(Z) < 3$ is moderate evidence in support of $M_{2}$ over $M_{1}$, while $0 < \Delta\ln(Z) < 1$ is inconclusive. For the sake of comparison to literature we have also included reduced $\chi^{2}$ values for model selection assessment.

Note that each flux density measurement will have a systematic uncertainty associated with the absolute flux density scale, which could cause the ensemble of measurements to deviate from a Gaussian distribution. To test whether this would have an impact on the fitting process we introduced hyperparameters to the fits \citep{Hobson2002}. Hyperparameters are similar to nuisance parameters in that they are marginalized over when calculating the posterior distribution, but hyperparameters differ in that they are not present in the model beforehand. They provide a method to quantify inaccurate uncertainties and systematic errors via weighting of the data sets \citep{Hobson2002}. We find no statistical evidence that a hyperparameter is justified to model a possible deviation from a normal distribution for the ensemble of measurements, with $\Delta\ln(Z) < 1$ when comparing fits with and without a hyperparameter. The reason the ensemble distribution closely adheres to a Gaussian distribution is likely due to the fact that PKS~B1934-638 was used to set the flux density scale for the majority of the observations.

\section{Results}\label{results}

The different absorption models for radio galaxies can be separated into two broad categories based on the underlying absorption mechanism: FFA and SSA. For FFA, the models vary depending on whether the ionized screen is internal or external to the radio emitting plasma, and on the topology of the screens. Additionally, the spectrum can also exhibit breaks due to the ageing of the electron population. In this section we first justify the use of multi-epoch data to fit the spectra of \pkssrcnospace, and then outline the different features of the absorption models and fitting statistics.

\subsection{Variability}\label{variability}

When considering the use of multi-epoch data to describe the spectrum of \pkssrcnospace, it is important to determine whether \pkssrc displays any temporal variability that could result in deviations to its intrinsic spectrum. \pkssrc has been part of several radio monitoring campaigns, since it is used as a primary or secondary calibrator for many southern hemisphere radio telescopes. \citet{CampellWilson1994} and \citet{Gaensler2000} both studied \pkssrc with the MOST to ascertain whether it displayed any evidence of intrinsic variability or scintillation at 0.843\,GHz. \pkssrc was observed irregularly over 22,000 times in a 2-minute scan mode between 1987 and 1996, with separation between the observations ranging from several hours to days. In order to test for variability, \citet{Gaensler2000} calculated the $\chi^{2}$ probability that the flux density has remained constant from the light curve of \pkssrcnospace, binned in 30 day intervals. This bin size was selected to reduce the effects of unrecognized systematic errors in the flux density determination and the presence of confusing sources in the field. \pkssrc was found to be non-variable on this time scale following the procedure of \citet{Kesteven1976}, which compares the probability of whether the dispersion in the weighted mean flux density comes from intrinsic variability or distributions of measurement errors, with a $p$-value significantly greater than 0.01. 

\pkssrc was also part of a monitoring campaign by the Tasmanian Mount Pleasant 26\,m antenna between 1990 and 1993. \pkssrc was regularly observed over a 30 month period 52 and 34 times at 2.3 and 8.4\,GHz, respectively \citep{Jauncey2003}. The average root-mean-square errors of the observations were $\sim$\,0.1\,Jy. Using the same statistical technique applied by \citet{Gaensler2000}, \citet{King1994} demonstrated that \pkssrc was not variable at 2.3\,GHz and 8.4\,GHz with $p$-values of 0.02 and 0.95, respectively. 

Therefore, it is evident that \pkssrc is not variable, within the uncertainties of these datasets, over timescales from hours to decades at several frequencies. This lack of variability is consistent with the nature of most GPS sources that have morphologies consistent with CSOs \citep{Odea1998,2001AJ....122.1661F}, and justifies the use of multi-epoch data to model the spectrum.
                                     
\subsection{Free-Free Absorption}\label{sect:free-free}

FFA involves the attenuation of emission radiation by an external or internal ionized screen relative to the emitting electrons. The morphology of the screen could be either homogeneous or inhomogeneous. For electron temperature $T_{\mathrm{e}}$ in K and free electron density $n_{\mathrm{e}}$ in cm$^{-3}$, the optical depth due to FFA at frequency $\nu$ in GHz is approximated by

\begin{equation}\label{opticaldepth}
	\tau_{\nu} \approx 8.24 \times 10^{-2} \nu^{-2.1} T_{\mathrm{e}}^{-1.35} \int n^{2}_{\mathrm{e}}~\mathrm{d}l,
\end{equation} 

\noindent where $l$ is the distance through the source in pc \citep{Mezger1967}. Assuming that a homogeneous ionized screen surrounds the plasma producing the non-thermal power-law spectrum, the resulting free-free absorbed spectrum is described by

\begin{equation}\label{singhomobremss}
	S_{\nu} = a \nu^{\alpha} e^{-\tau_{\nu}},
\end{equation}

\noindent where $a$ characterizes the amplitude of the intrinsic synchrotron spectrum, and $\alpha$ is the spectral index of the synchrotron spectrum. It is convenient to parameterize the optical depth as $\tau_{\nu} = (\nu/\nu_{\mathrm{p}})^{-2.1}$, where $\nu_{\mathrm{p}}$ is the frequency at which the optical depth is unity. The best model fit for the spectrum is shown in Figure \ref{sedfigs}(a) and the resulting parameter values, with associated 1-$\sigma$ uncertainties, are presented in Table \ref{longtable}.

As is clear from the fitting statistics, a single homogeneous absorbing screen cannot accurately describe the spectrum of \pkssrcnospace: the gradient below the turnover is too steep and it overpredicts the flux density at high frequencies. One possible alternative, as suggested by \citet{Tingay2003}, is to fit two separate homogeneous free-free absorbing screens in front of two separate non-thermal distributions of electrons, since the source has been resolved into two components. The justification for this method is that for all observations with lower resolution than VLBI, the flux density we measure represents the sum of the environments of both these components. Therefore, the double homogeneous free-free model fit to the spectrum has the form:

\begin{equation}\label{doubhomobremss}
	S_{\nu} = \sum_{i=1,2} a_{i} \nu^{\alpha_{i}} e^{-\tau_{\nu,i}}.
\end{equation}

\noindent The resulting fit is also presented in Figure \ref{sedfigs}(a) and Table \ref{longtable}. The double homogeneous FFA model is a statistical improvement over the single homogeneous model, with $\Delta\ln(Z) > 6$. However, despite doubling the number of parameters in the model, the double homogeneous FFA model still falls too rapidly at low radio frequencies and over predicts the flux density at high frequencies.

It is possible that the absorbing ionized medium is mixed in with the relativistic electrons that are producing the non-thermal spectrum. In that case the spectrum is characterized as

\begin{equation}\label{internalbremss}
	S_{\nu} = a\nu^{\alpha}\left(\dfrac{1-e^{-\tau_{\nu}}}{\tau_{\nu}}\right).
\end{equation}

\noindent This fit is presented in Figure \ref{sedfigs}(b) and in Table \ref{longtable}. It is evident that the internal FFA cannot accurately describe the spectrum because of its shallow slope below the turnover. This is to be expected since the gradient below the turnover for such a model should be $\alpha + 2.1$.

An alternative FFA model, proposed by \citet{Bicknell1997}, treats the FFA screen as inhomogeneous and external to the non-thermal electrons in the lobes of the source. In this model, the jets from the AGN produce a bow shock that photoionizes the interstellar medium as it propagates, such that if the density of the gas surrounding the radio lobe is sufficiently high, it will cause FFA. Hence, the clouds have a range of optical depths that \citet{Bicknell1997} assume follow a power-law distribution $\eta$ parameterized by the index $p$, such that $\eta^{p} \propto \int (n_{\mathrm{e}}^{2}T_{\mathrm{e}}^{-1.35})^{p} \mathrm{d}l$. Note that we require $p > -1$ otherwise the model reduces to the homogeneous limit, and $\eta$ varies from 0 to a maximum value $\eta_{0} = \nu_{\mathrm{p}}^{-2.1}$. Assuming that the length scale of the shock and the inhomogeneities in the interstellar medium are both much less than the size of the lobes, the model can be represented by

\begin{equation}\label{singinhomobremss}
	S_{\nu} = a(p+1)\gamma\left[p+1,\left(\dfrac{\nu}{\nu_{\mathrm{p}}}\right)^{-2.1}\right]\left(\dfrac{\nu}{\nu_{\mathrm{p}}}\right)^{2.1(p+1)+\alpha} ,
\end{equation}

\noindent where $\gamma$ is the lower incomplete gamma function of order $p + 1$. The model is plotted in Figure \ref{sedfigs}(c) and the fit parameters are represented in Table \ref{longtable}. This is the best fitting FFA model with $\Delta\ln(Z) > 3$ when compared to the double homogeneous model. Comparatively to the double homogeneous FFA model, the inhomogeneous FFA model provides a better fit to the slope below the turnover but continues to predict additional flux at high frequencies. 

\begin{sidewaystable}
	\renewcommand{\arraystretch}{2}
	\scriptsize
	\centering
	\caption{\label{longtable} Best fit parameters and their associated uncertainties for the different absorption models. The uncertainties presented are 1-$\sigma$. Note the reported values are for the observed frame of \pkssrcnospace. Free-free absorption and synchrotron self-absorption are abbreviated to FFA and SSA, respectively. The parameters listed in the table are: the normalization parameters of the intrinsic synchrotron spectrum $a_{i}$, the spectral indices of the synchrotron spectrum $\alpha_{i}$, the power-law indices of the electron energy distribution $\beta_{i}$, the frequencies at which the optical depth of the absorption model is unity $\nu_{\mathrm{p},i}$, the high-frequency exponential cutoff $\nu_{\mathrm{br}}$, the reduced $\chi^{2}$-value of the model fit $\chi^{2}_{\mathrm{red}}$, and the log of the Bayesian evidence of the model fit $\ln(Z)$.}
		\begin{tabular}{lcccccccccccc}
		\hline
		\hline
Models & $a_{1}$ (Jy) &  $a_{2}$ (Jy) & $\alpha_{1}$ & $\alpha_{2}$ & $\beta_{1}$ & $\beta_{2}$  & $\nu_{\mathrm{p},1}$ (GHz)  & $\nu_{\mathrm{p},2}$ (GHz)  & $p$  & $\nu_{\mathrm{br}}$ (GHz) & $\chi^{2}_{\mathrm{red}}$ & $\ln(Z)$ \\
				\hline		
	Single Homogeneous FFA  & 7.99 $\pm$ 0.08 &  $\cdots$ &  $-1.190$ $\pm$ 0.006    & $\cdots$  &  $\cdots$	& $\cdots$ & 0.480 $\pm$ 0.007   &  $\cdots$ & $\cdots$ &$\cdots$   & 12.18 & $-567.4$ $\pm$ 0.2\\
	 Double Homogeneous FFA & 3.6 $\pm$ 0.2 &  5.8 $\pm$ 0.2 & $-1.27^{+0.05}_{-0.04}$  &  $-1.28$ $\pm$ 0.03 & $\cdots$	& $\cdots$  & 0.293 $\pm$ 0.007  & 0.93 $\pm$ 0.05  & $\cdots$ & $\cdots$ & 5.52 & $-422.4$ $\pm$ 0.2 \\
	 Inhomogeneous FFA  & 12.0 $\pm$  0.3 &   $\cdots$ &  $-1.221$ $\pm$ 0.008 & $\cdots$  &$\cdots$ &$\cdots$ & 0.75 $\pm$ 0.02 & $\cdots$ & 0.24 $\pm$ 0.04 & $\cdots$ & 5.72& $-419.0$ $\pm$ 0.6\\
	 Internal FFA & 8.8 $\pm$ 0.1 &  $\cdots$ & $-1.240$ $\pm$ 0.008  &  $\cdots$ & $\cdots$& $\cdots$ & 0.90 $\pm$ 0.02 & $\cdots$ & $\cdots$ & $\cdots$ & 7.13 & $-567.4$ $\pm$ 0.2\\
	   Single SSA & 13.4 $\pm$ 0.2  & $\cdots$ & $\cdots$ &$\cdots$ & 3.27 $\pm$ 0.01 	& $\cdots$ & 0.575 $\pm$ 0.008  & $\cdots$   & $\cdots$ & $\cdots$ & 13.83& $-609.5$ $\pm$ 0.1 \\
	Double SSA & 12.2 $\pm$ 0.3 & 2.5 $\pm$ 0.2 & $\cdots$ & $\cdots$ & 3.48 $\pm$ 0.06 &  3.56$^{+0.08}_{-0.07}$ & 0.459 $\pm$ 0.009  &  1.56 $\pm$ 0.7 & $\cdots$& $\cdots$ &5.36 & $-408.5$ $\pm$ 0.3\\
	 Single SSA + exp. break & 12.4 $\pm$  0.2 	& $\cdots$ &$\cdots$ &  $\cdots$ & 2.61 $\pm$ 0.03 & $\cdots$  & 0.75 $\pm$ 0.02  & $\cdots$ & 13.7 $\pm$ 0.7 & $\cdots$ & 2.39 & $-336.6$ $\pm$ 0.2\\
	 Double SSA + exp. break & 8.3$^{+0.6}_{-0.7}$ & 6.1$^{+0.7}_{-0.6}$  & $\cdots$  &  $\cdots$ & 3.3$^{+0.3}_{-0.2}$ &  2.55$^{+0.06}_{-0.08}$ & 0.34$^{+0.07}_{-0.08}$   &  0.73$^{+0.04}_{-0.05}$ & $\cdots$ & 12 $\pm$ 2 & 1.09 & $-305.8$ $\pm$ 0.3 \\
	 Inhomogeneous FFA + exp. break & 12.9 $\pm$  0.3	& $\cdots$ & $-0.93$ $\pm$ 0.02  & $\cdots$ &$\cdots$ & $\cdots$& 0.55 $\pm$ 0.02  & $\cdots$ & 0.37$^{+0.07}_{-0.06}$ & 19 $\pm$ 1 & 0.80 & $-304.1$ $\pm$ 0.1\\
		\hline
		\end{tabular}
\end{sidewaystable}

\subsection{Synchrotron Self-Absorption}

SSA has been the other dominant theory used to explain the inverted spectra of GPS sources. In this model, the turnover occurs because the source cannot have a brightness temperature that exceeds the plasma temperature of the non-thermal electrons \citep{Kellermann1966}. Alternatively, SSA can be thought to occur at the frequency at which the relativistic electrons and emitted synchrotron photons have a high chance of scattering. The absorption cross-section for a synchrotron electron and a low-energy photon is greater at larger wavelengths. Hence, for a source of a given size, very-long wavelength emission is only visible from a very thin shell at the surface of the source. As the observing frequency increases, photons emerge from regions of the source that are progressively deeper in the source such that the total flux density increases until the optically thin regime is reached. Parameterizing the spectrum in terms of the power-law index, $\beta$, of the electron energy distribution \citep{Tingay2003}, such that $\alpha = -(\beta - 1) /2$, and assuming the synchrotron emitting region is homogeneous, the spectrum is modeled by:

\begin{subequations}
\begin{eqnarray}\label{singSSAeqn}
	S_{\nu} = a\left(\dfrac{\nu}{\nu_{\mathrm{p}}}\right)^{-(\beta - 1)/2}\left(\dfrac{1 - e^{-\tau}}{\tau}\right),
\end{eqnarray}
\noindent where,
\begin{eqnarray}
 \tau = \left(\dfrac{\nu}{\nu_{\mathrm{p}}}\right)^{-(\beta + 4)/2},
\end{eqnarray}
\end{subequations}

\noindent and where $\nu_{\mathrm{p}}$ corresponds to the frequency at which the source becomes optically thick. This is the frequency at which the mean free path of electron-photon scatterings is approximately the size of the source. Applying the same logic as in \S\,\ref{sect:free-free}, that the observed flux density from the source is the sum of two different lobes with different electron energy distributions, the spectrum could also be described by two SSA components such that
 
\begin{subequations}
\begin{eqnarray}\label{doubSSAeqn}
	S_{\nu} = \sum_{i=1,2} a_{i}\left(\dfrac{\nu}{\nu_{\mathrm{p},i}}\right)^{-(\beta_{i} - 1)/2}\left(\dfrac{1 - e^{-\tau_{i}}}{\tau_{i}}\right),
\end{eqnarray}
\noindent where,
\begin{eqnarray}
 \tau_{i} = \left(\dfrac{\nu}{\nu_{\mathrm{p},i}}\right)^{-(\beta_{i} + 4)/2}.
\end{eqnarray}
\end{subequations}

\begin{figure*}
\begin{center}$
\begin{array}{cccc}
\includegraphics[scale=0.27]{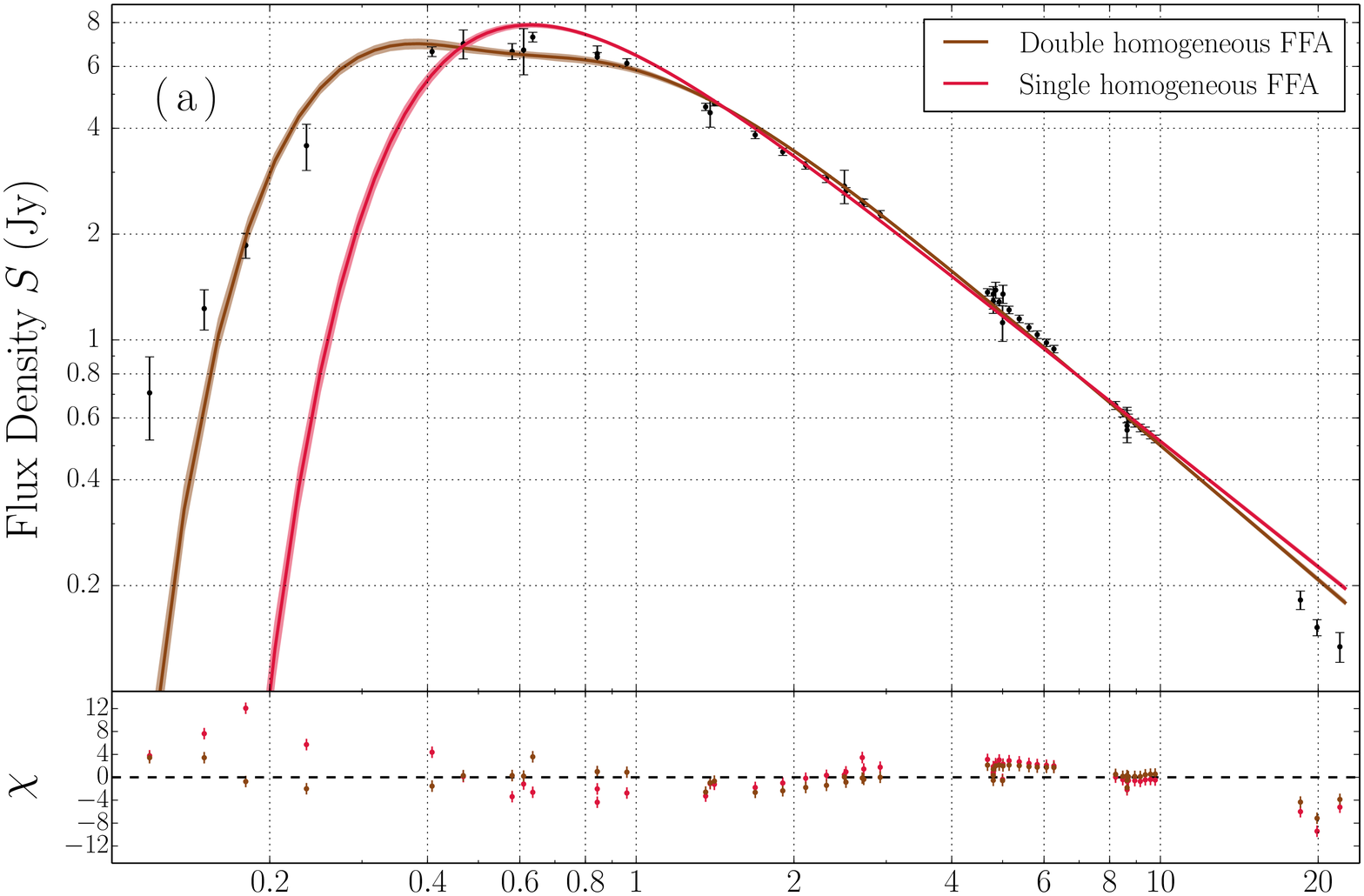} &
\includegraphics[scale=0.27]{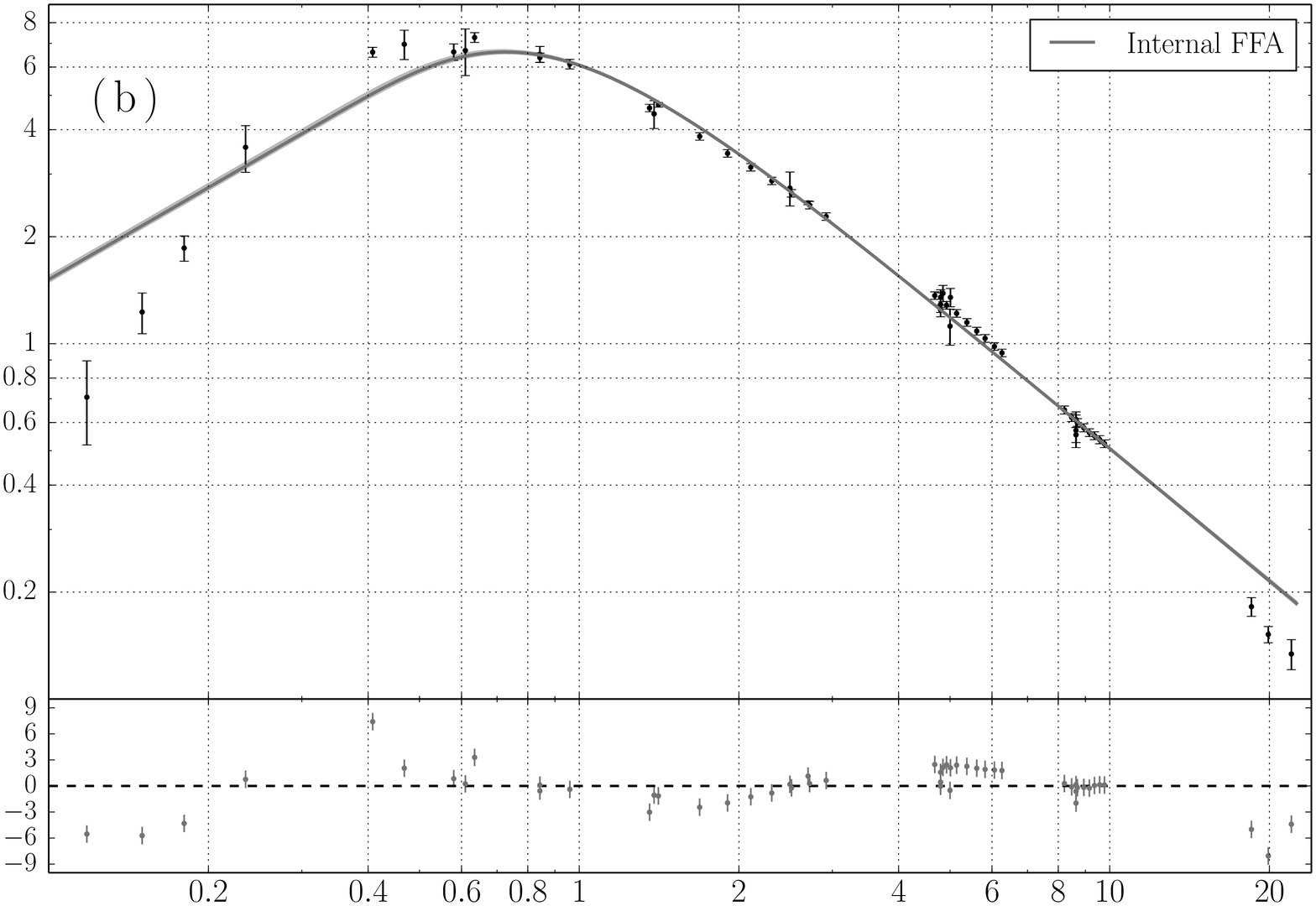}\\
\includegraphics[scale=0.27]{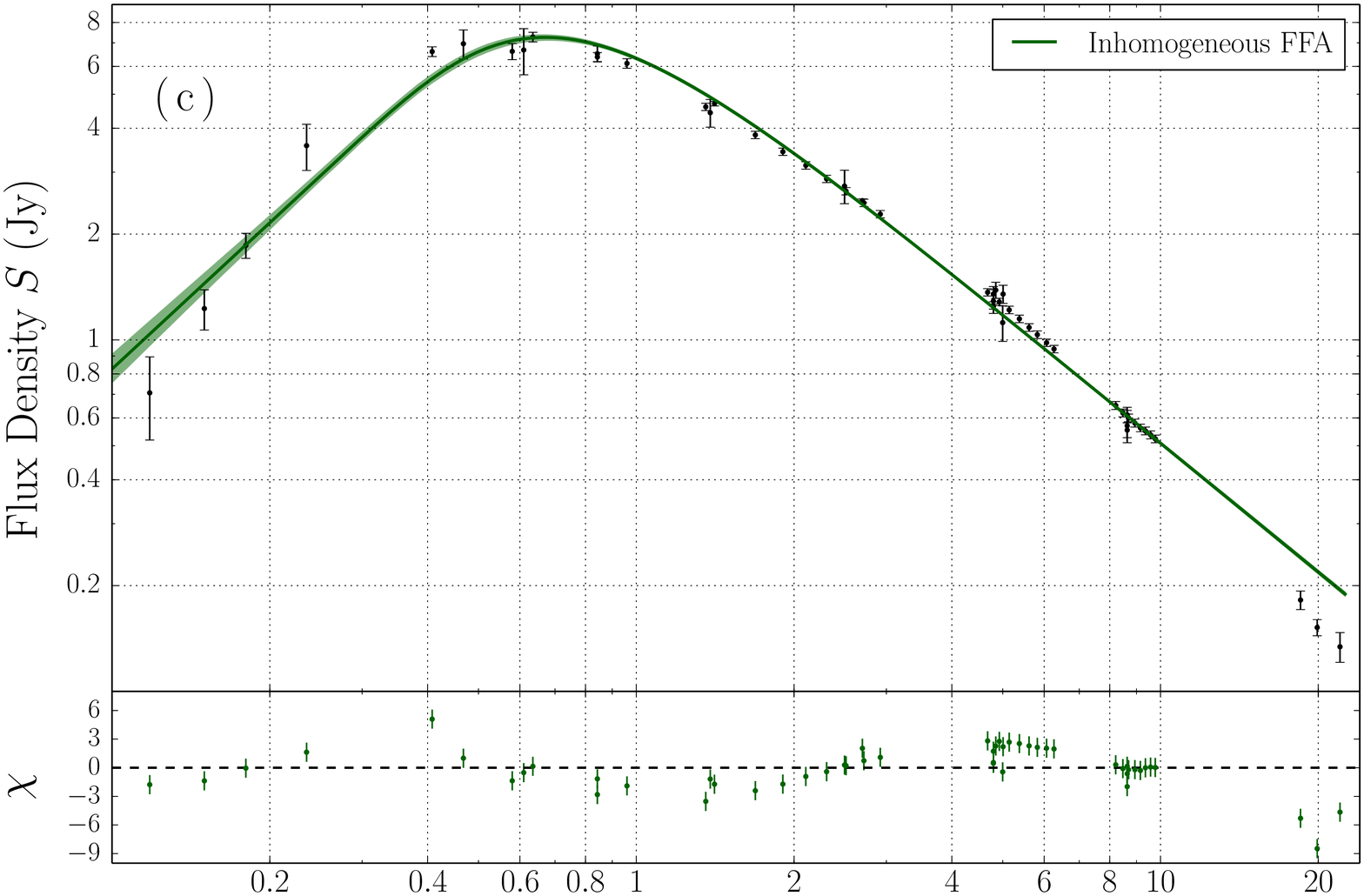} &
\includegraphics[scale=0.27]{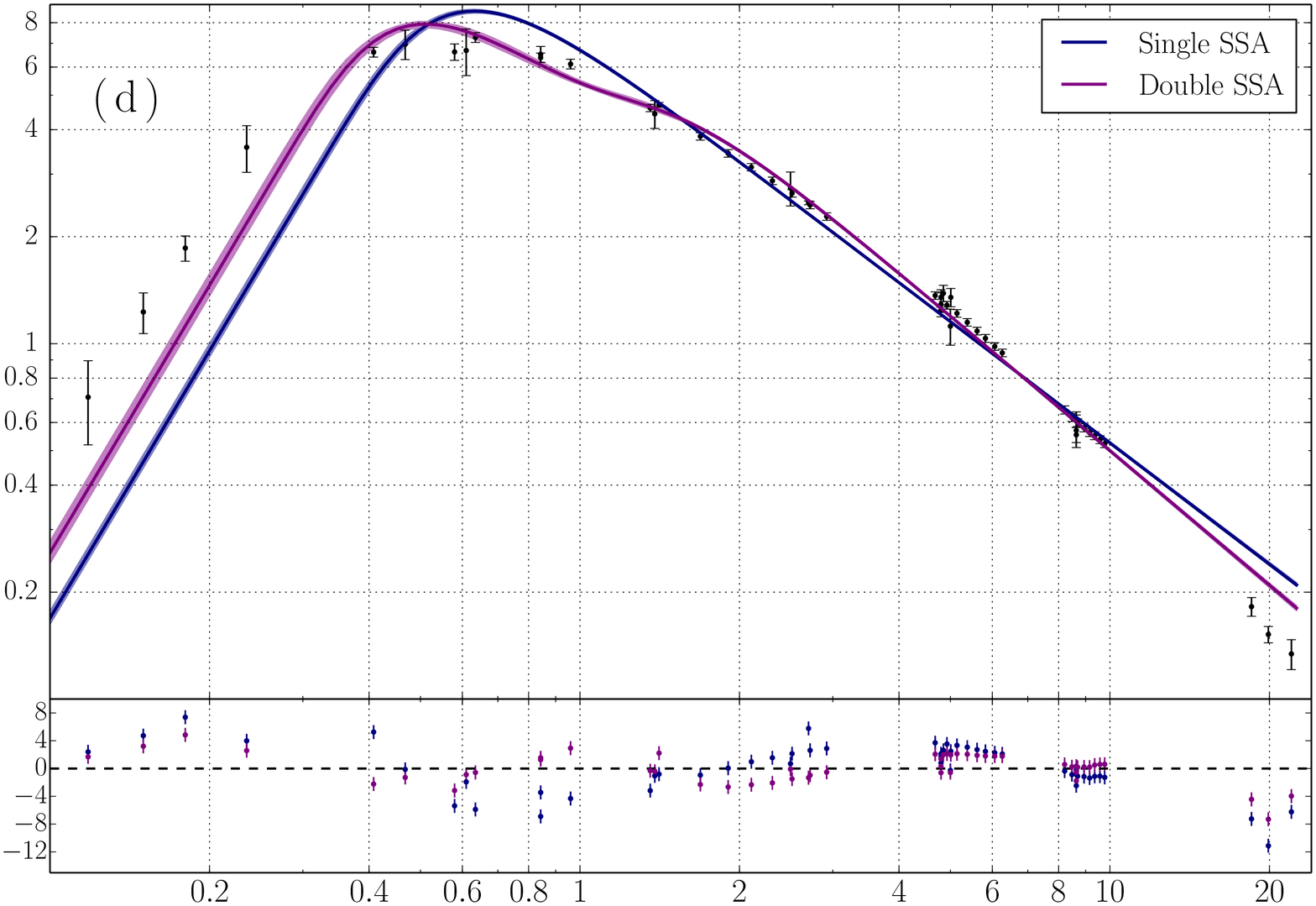}\\
\includegraphics[scale=0.27]{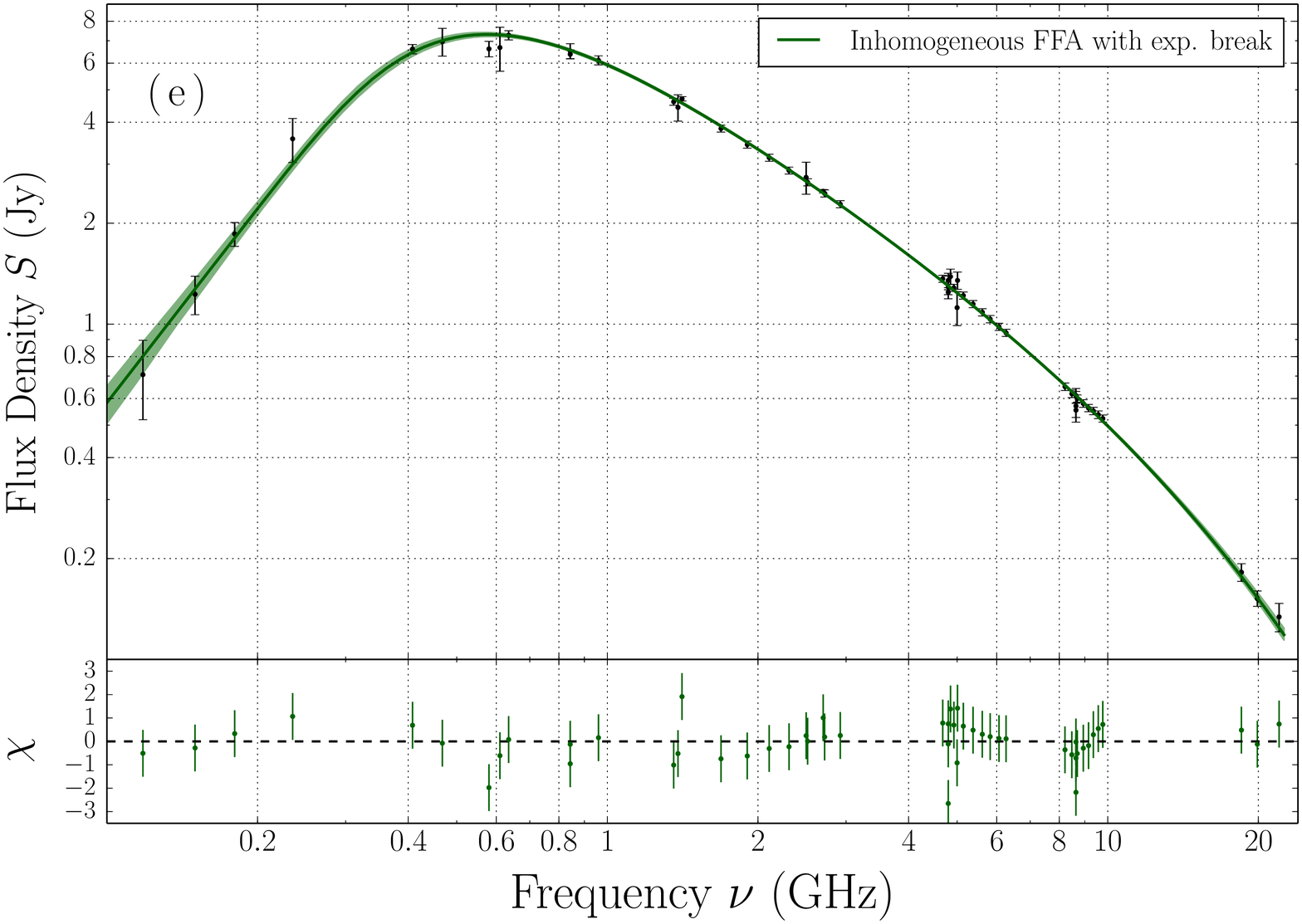} &
\includegraphics[scale=0.27]{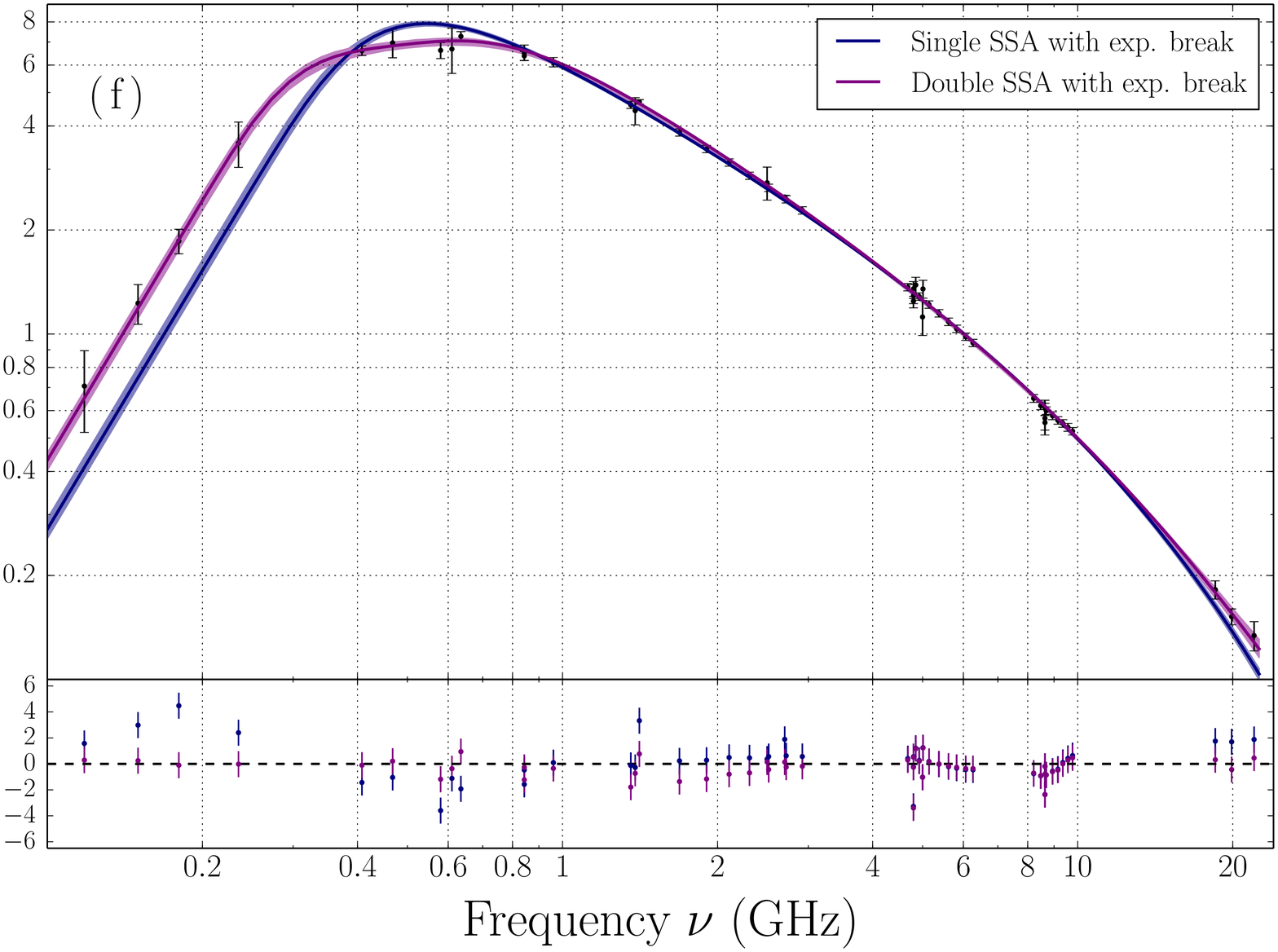}
\end{array}$
 \caption{Different absorption model fits to the spectral energy distribution of \pkssrcnospace, using the parameter values reported in Table \ref{longtable}. The colored regions associated with the models represent the 1-$\sigma$ uncertainty on the model fit at the respective frequency. The $\chi$-values for the model fits are displayed below the spectral energy distributions. Note that free-free absorption and synchrotron self-absorption in these figures are abbreviated to FFA and SSA, respectively. (a) Single and double homogeneous FFA models are shown in brown and red, respectively. (b) Internal FFA model is represented in gray. (c) Inhomogeneous FFA model is shown in green. (d) Single and double SSA model fits are presented in purple and blue, respectively. (e) Inhomogeneous FFA model with an exponential spectral break is shown in green. (f) Single and double SSA models, with exponential spectral breaks, are presented in purple and blue, respectively. All fits are conducted in the observed frame.}
\label{sedfigs}
\end{center}
\end{figure*}

\noindent The fits for the single and double SSA models are presented in Table \ref{longtable} and Figure \ref{sedfigs}(d). The double SSA model is statistically better fit than the single SSA model, with $\Delta\ln(Z) > 100$. However, as evident in Figure \ref{sedfigs}(d), both models accurately model the slope below the turnover but underestimate the flux density at low frequencies and overestimate it at high frequencies.

We do not outline an inhomogeneous model of SSA since the slope below the turnover in the spectrum of \pkssrc is close to the theoretical prediction of the homogeneous SSA model. An attempt to fit a more general inhomogeneous SSA model finds that the steep spectral slope forces the model fit to be degenerate with the homogeneous SSA model. 

\subsection{Spectral breaks and cut-offs}

As the non-thermal electrons age in the jets and lobes of a radio source, synchrotron and inverse-Compton losses preferentially deplete high-energy electron populations such that the spectrum deviates from a single power-law. There are several models that predict how the shape of the spectrum should evolve. In the continuous injection model of \citet{Kardashev1962} a stream of electrons is constantly injected into a volume permeated with a constant magnetic field. This model predicts that the optically thin part of the spectrum should steepen above a break frequency, $\nu'_{\mathrm{br}}$, from $\alpha$ to $\alpha - 0.5$. Once this characteristic break frequency is known, the source's age and magnetic field strength can be estimated \citep{Murgia1999}. We find that augmenting the models in the previous section with the continuous injection model does not improve the fit to the data because of its discontinuous nature. The abrupt change in spectral index is inconsistent with the smooth, continuous decline evident in the three bands of CABB data in Figure \ref{nomodelsed}, and the fact the decrease in slope is not steep enough to fit this decline.

For GPS sources the assumptions of the continuous injection model may not hold, since there is evidence of intermittent and recurrent activity in the nuclei of such sources \citep[e.g.][]{Owsianik1998,Brocksopp2007}. Approximately ten percent of GPS sources display several large scale radio structures that are usually explained as relics of previous active phases \citep{Stanghellini2005}. An alternative model is to assume that the injection of fresh particles has ceased and that the radio source is already in the relic phase. In addition to the transition to the steeper slope $\alpha - 0.5$ after $\nu'_{\mathrm{br}}$, this model predicts the formation of a high-frequency exponential cutoff after a second break frequency, $\nu_{\mathrm{br}}$, such that the absorption models have an additional multiplicative factor $e^{-\nu/\nu_{\mathrm{br}}}$ \citep{Jaffe1973,Komissarov1994,Murgia2003}. This second break frequency is related to the first according to

\begin{equation}\label{expbreakeqn}
 \nu_{\mathrm{br}} = \nu'_{\mathrm{br}}\left(\dfrac{t_{\mathrm{s}}}{t_{\mathrm{off}}}\right)^{2},
\end{equation}

\noindent where $t_{\mathrm{s}}$ is the age of the source and $t_{\mathrm{off}}$ is the time since the injection of fresh electrons has ceased. 

As mentioned above, a steepening of the optically thin slope to $\alpha - 0.5$ is not statistically evident for \pkssrcnospace. However, it is possible that the first break frequency $\nu'_{\mathrm{br}}$ has moved to a lower frequency below the spectral turnover. This is consistent with the fact that the absorption models require an extraordinarily steep injection spectral index of $\alpha \sim$\,$-1.2$ to fit the high-frequency data of \pkssrcnospace. If the injection has ceased and the first spectral break frequency has moved to the optically thick part of the spectrum, a more reasonable injection spectral index of $\alpha \sim$\,$-0.7$ can fit the optically thin part of the spectrum. As is clear in Table \ref{longtable}, the addition of an exponential attenuation at high frequencies provides a very significant statistical improvement in the overall fit to the spectrum, with $\Delta\ln(Z) > 100$ for corresponding models compared with and without the exponential break. Note that while all the models are improved by the exponential break, only the three best fitting models are presented in Figures \ref{sedfigs}(e) and \ref{sedfigs}(f) as the addition of the exponential break still does not allow the spectral slope below the turnover to be well fit for homogeneous free-free and internal FFA models. After the addition of the high-frequency exponential break, the two best fitting models are the double SSA and the inhomogeneous FFA. The difference in Bayesian evidence between these two models, $\Delta\ln(Z) = 1.7$, moderately favors the inhomogeneous FFA model with an exponential break over the double SSA model with an exponential break. Such exponential spectral breaks have been observed in other sources, for instance PKS B1518+047 \citep{Orienti2010}. Significantly, not only does this addition allow the models to follow the curvature evident in the CABB bands, it also ensures that the double SSA and inhomogeneous FFA models do not underestimate the flux density at low frequencies. This indicates that it is necessary to have good high-frequency coverage to correctly model the flux below the turnover in the spectrum. 

Note that the high-frequency steepening could be due to source structure gradually being resolved out when observing from low to high frequencies or by the presence of older extended structure. As discussed in  \S\,\ref{datared}, the steepening cannot be due to resolution affects since \pkssrc was unresolved at all frequencies, even when 6\,km baselines are used at 20\,GHz \citep{Chhetri2013}. All imaging at the frequencies of the ATCA observations, and at VLBI resolutions, also did not provide any evidence for extended structure. Therefore, the break evident in the spectrum is highly likely to be a physical feature of \pkssrcnospace. 

\section{Discussion}\label{discussion}

\begin{figure}
\centering
\includegraphics[scale = 0.265]{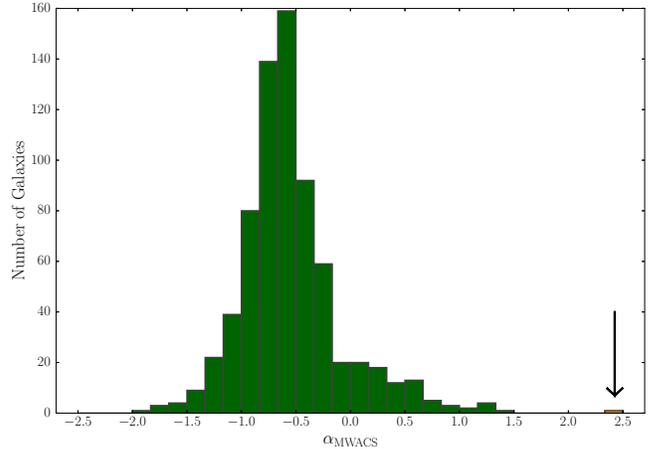}
\caption{\label{dist_alpha} Distribution of the MWACS spectral index for the 706 sources detected in both MWACS and AT20G. Such a sample is dominated by compact sources. $\alpha$ in this plot represents the spectral index of the source in the MWACS observations only. The location of \pkssrc is highlighted in orange and by an arrow. This location is over five standard deviations away from the median of the distribution.}
\end{figure}

\pkssrc represents an extreme outlier in the population of GPS sources. It has the smallest spectral width (defined as the fullwidth at half maximum), $\sim$\,0.5 decade of frequency, and steepest low radio frequency slope, $\alpha \sim$\,$+2.4$, of any known GPS source. The average spectral width and spectral index below the turnover for GPS sources are $\sim$\,1.2 decades of frequency \citep{Odea1991} and $\alpha \sim$\,$+0.7$ \citep{deVries1997}, respectively. The outlier nature of \pkssrc is emphasized in Figure \ref{dist_alpha}, which shows the distribution of the spectral indices of MWACS sources that have a AT20G counterpart. Such a population is dominated by bright, compact radio sources, likely including many GPS sources. The spectral slope of \pkssrc in the MWACS band is over five standard deviations away from the median of the distribution.

Since \pkssrc has an observed peak frequency below a gigahertz and a double component morphology that has a physical separation of $\sim$\,1000\,pc, it has several similarities to a CSS source. However, when the relationships between the observed spectral peak, radio power, angular size and peak brightness are compared (e.g. Fig.~1, \citealt{2000MNRAS.319..445S}; Fig.~1, \citealt{An2012}), all of PKS B0008-421 spectral properties are consistent with the GPS/CSO classification. Additionally, the physical separation between the double components is too large to for \pkssrc to be associated with the HFP population \citep{2006A&A...450..959O}.

It is noteworthy that the slope of the spectrum below the turnover of \pkssrc is close to the SSA theoretical limit of $\alpha =$\,$+2.5$. This is because the common explanation for observing shallower spectral gradients below the turnover is due to inhomogeneity of the SSA regions in the source \citep{Odea1998,Tingay2003}. Such a steep low-frequency gradient could be indicative of a homogeneous emitting structure that could allow the most comprehensive differentiation between the different types of absorption models. Additionally, the very steep optically thin spectral index of $\alpha \sim$\,$-1.2$ is a surprising characteristic since, if the youth scenario is correct, the radio emission from GPS sources should have only started several thousand years ago. This would imply an extraordinary injection spectral index is needed, or a source significantly older than common for GPS sources.

As shown in \S\,\ref{results}, we can statistically conclude that internal FFA, single and double homogeneous FFA, and single SSA cannot be responsible for the turnover in the spectrum of \pkssrcnospace. The two models that best fit the spectrum of \pkssrc are the inhomogeneous FFA and double SSA models, both with the addition of a high-frequency exponential break. Both models can accurately predict the flux density and slope of the spectrum at high and low radio frequencies. By eye the two fits are identical but the difference in Bayesian evidence, $\Delta\ln(Z) = 1.7$, moderately favors the inhomogeneous FFA model over the double SSA model. 

However, both the inhomogeneous FFA and double SSA models require some abnormal physical parameters to generate their fit, underscoring the extreme nature of \pkssrcnospace. Using the best fitting parameters from the double SSA model with an exponential break, the magnetic field strength of the source can be estimated. Provided that the spectral turnover of GPS sources is caused by SSA, the turnover frequency $\nu_{\mathrm{max}}$, in GHz, for a homogeneous, self-absorbed radio source with a power-law electron energy distribution can be expressed \citep{Kellermann1981} in terms of the magnetic field strength $B$, in G, the flux density at the turnover frequency $S_{\mathrm{max}}$, in\,Jy, and angular size of the source $\theta_{\mathrm{max}}$, in milliarcseconds, by

\begin{equation}\label{turnoverSSA}
 \nu_{\mathrm{max}} \sim 8.1B^{1/5}S_{\mathrm{max}}^{2/5}\theta^{-4/5}(1+z)^{1/5}.
\end{equation}

\noindent For \pkssrcnospace, this is estimated to be $B \sim\,$$4.1$ G. This is an order of magnitude larger than the typical magnetic field strength reported for GPS sources, which are often in the range of 5 to 100 mG \citep{Odea1998,Orienti2008}. 

In comparison, the inhomogeneous FFA model demands a large, positive $p$ value to fit the steep low-frequency spectral slope, as shown in Table \ref{longtable}. All previous literature values report values of $p$ that are negative or zero because of the more common shallow slope below the turnover \citep{Bicknell1997,Tingay2003}. Since $p$ parameterizes the distribution of optical depth of absorbing clouds, the implication of a large, positive $p$ is that there are substantially more clouds of high emission measure than clouds of low emission measure. In terms of the environment of the source, this could be justified if the radio jet is beginning to propagate into the environment and breaking up the interstellar medium into several large, dense and cool clouds. Therefore, a positive $p$ is consistent with the idea that the source is confined to a small spatial scale due to high density, low temperature clouds. If the jets continue to propagate into the medium, the jet interactions with the surrounding medium will increase the temperature and decrease the density of the clouds, as evident in \citet{Wagner2011}, such that $p$ will decrease towards more common literature values.  

Independent of the underlying absorption model, the necessity of a high-frequency exponential break to accurately fit the spectrum means that injection or acceleration of relativistic particles has ceased in the source \citep{Orienti2010}. The lack of variability also suggests that \pkssrc is no longer being powered. The presence of a high energy exponential break, but no statistical evidence of the first spectral break frequency, indicates that the first break has moved to the optically thick part of the spectrum, $\nu'_{\mathrm{br}} \lesssim 0.59$\,GHz. This allows a more typical injection spectral index $\alpha \sim$\,$-0.7$ to fit the optically thin part of the spectrum. With the knowledge of the spectral break frequencies, we can place a lower limit on the age the source $t_{\mathrm{s}}$ and calculate the time since the last injection of electrons $t_{\mathrm{off}}$ using 

\begin{equation}\label{breakage}
 t_{\mathrm{s}} = 5.03 \times 10^{4}B^{-3/2}\nu'_{\mathrm{br}}(1+z)^{1/2}~\mathrm{yr},
\end{equation}

\noindent and Equation \ref{expbreakeqn}, where $B$ is in mG and the break frequencies are in GHz. 

We can estimate the magnetic field strength of \pkssrc independently of SSA through equipartition theory, assuming that the total energy densities of cosmic rays and the magnetic fields are equal. Using the revised formula of \citet{Beck2005}, the equipartition magnetic field for an injection spectral index of $\alpha = -0.7$ is $B_{\mathrm{eq}}$ $\sim$\,6.2 mG. This magnetic field strength is inconsistent with that derived from SSA theory, but consistent with other magnetic field strengths derived for GPS sources \citep{Orienti2008}. Using the equipartition magnetic field strength, the radiative lifetime of \pkssrc is calculated to be $t_{\mathrm{s}} \gtrsim 2900$ years and, using $\nu_{\mathrm{br}}$ from the inhomogeneous FFA model, the time since the last injection of fresh electrons is $t_{\mathrm{off}} = 550 \pm 110$ years. Both of these values are consistent with the youth hypothesis of GPS sources \citep{Murgia1999}. Applying the much larger magnetic field strength calculated from SSA theory and the fitted $\nu_{\mathrm{br}}$ from the double SSA model, the radiative age lower limit is unrestrictive and we find $t_{\mathrm{off}}$ $\sim$\,$650,000$ years. This time period is inconsistent with a source with such a large flux density and provides a physical argument in favor of inhomogeneous FFA over the double SSA model.

Further evidence that the inhomogeneous FFA model is responsible for the turnover in the spectrum of \pkssrc is that the peak frequency occurs at around a gigahertz. If the injection of relativistic electrons has finished, strong adiabatic cooling should see a shift of the peak frequency quickly out of the gigahertz regime. Assuming the magnetic field is frozen into the plasma, the spectral peak would decrease from $\nu_{\mathrm{p},0}$ at time $t_{0}$ to $\nu_{\mathrm{p},1}$ at time $t_{1}$ with $\nu_{\mathrm{p},1} = \nu_{\mathrm{p},0}(t_{0}/t_{1})^{4}$ \citep{Orienti2008}. This would suggest the detection of a GPS source that has ceased actively fueling radio jets would be rare because the peak would quickly shift out of common observable radio frequencies. However, a possible way that \pkssrc has maintained its peak frequency close to a gigahertz, despite the cessation of fresh electrons around 500 years ago, is through the presence of a dense ambient medium. This dense medium would limit the adiabatic losses and maintain the observed peak brightness and frequency. This again favors the inhomogeneous FFA model to explain the spectrum of \pkssrc since the large, positive $p$ parameter needed for the fit also implies a dense surrounding medium. A test of this hypothesis would be to target \pkssrc for H\,\textsc{i} absorption studies and to determine whether the associated absorption towards the source has a column density $\gtrsim 10^{20}$\,cm$^{-2}$, since this model suggests it is probable much of the medium surrounding the jets is unionized \citep{Vermeulen2003,Tengstrand2009}.

It is important to emphasise that both the double SSA and inhomogeneous FFA model fits require extreme physical conditions to provide an adequate fit to the spectrum \pkssrcnospace. In the case of the double SSA, the model requires a magnetic field strength that is an order of magnitude larger than the average for the population of GPS sources. While for the case of FFA, the model requires a distribution of emission measure that is dominated by cold and dense clouds to provide an adequate fit. However, the magnetic field strength required by SSA is inconsistent with equipartition and with the characteristic time scales required by the exponential spectral break. Since the physical parameters required by FFA are independently consistent with the characteristic time scales of the exponential spectral break and with the fact that the peak of the spectrum is still near a gigahertz, we suggest that the extreme physical environment that the inhomogeneous FFA model requires is more likely than that required by the double SSA model fit.

The study of \pkssrc also has implications for the types of sources that will be discovered by the all-sky surveys being performed MWA and LOFAR. Many GPS sources that are at $z > 1$ and have ceased actively fueling their jets would not be identifiable at gigahertz wavelengths as classical GPS sources, since their spectral peaks would be below 0.3\,GHz. This means that these two telescopes could reveal high redshift GPS sources and a whole new population of GPS sources that have ceased activity and are fading. If a large number of GPS sources are identified as having ceased activity, it will be possible to conduct a comprehensive analysis as to whether such a population can explain the overabundance problem of CSS and GPS sources encountered by \citet{Readhead1996} and \citet{An2012}. \pkssrc also has implications for the identification of ultra-steep spectrum sources (USS), which are sources defined as having $\alpha < -1$. While \citet{Klamer2006} did not find high-frequency steepening for a sample of USS sources, \pkssrc would be identified as a USS source if no observations had been conducted below 0.6\,GHz. This could imply that at least part of the USS population consists of dying GPS sources where the spectral break has shifted below the low radio frequency turnover. Finally, the next generation of blind, large H\,\textsc{i} absorption studies, such as the planned H\,\textsc{i} absorption survey on the Australian Square Kilometre Array Pathfinder \citep[ASKAP;][]{Johnston2008}, will be vital in constraining the H\,\textsc{i} densities in GPS sources and could provide additional evidence for FFA.

\section{Conclusions}\label{conclusions}

We have presented broadband spectral modeling of the extreme GPS source \pkssrcnospace, which represents an outlier in the current GPS population since it has the steepest spectral slope below the turnover, and smallest spectral width, of any known GPS source. \pkssrc is part of the CSO morphological class, and its lack of variability allowed the use of multi-epoch data to describe the spectrum, with spectral coverage from 0.118 to 22\,GHz. In this analysis we have:

\begin{enumerate}
	
	\item{Determined the two best fitting spectral models are double SSA and inhomogeneous FFA. We also statistically excluded internal FFA, single and double component homogeneous FFA, and single component SSA as being explanations for the turnover in the spectrum of \pkssrcnospace.}
	
	\item{Demonstrated the necessity of a high-frequency exponential break to adequately describe the spectrum of the source. This implies that the source has a spectral break below the turnover and has likely ceased the injection of fresh particles. With the requirement of a high-frequency exponential break, the inhomogeneous FFA model is moderately statistically favored over the double SSA model.}
	
	\item{Highlighted that since activity in the source has ceased, constraints can be placed on the time since last injection. If the physical parameters derived from the double SSA fit are used, the time periods that result are unrealistic. However, parameters from the inhomogeneous FFA model provide a time since the last injection of fresh electrons $t_{\mathrm{off}} = 550 \pm 110$ years. The existence of a spectral peak in the gigahertz range requires a high density ambient medium to restrict adiabatic losses. This conclusion is congruous with the dense ambient medium required by the inhomogeneous FFA fit. These two lines of evidence lead us to conclude that inhomogeneous FFA is responsible for the turnover in the spectrum of \pkssrcnospace.}

\end{enumerate}

These conclusions imply it is likely that MWA and LOFAR will expose a new population of CSS and GPS sources that have ceased activity. Such sources would not have been identified from previous high-frequency surveys since the spectral peak would have shifted below a gigahertz. If such a population is found to be prominent, it could help explain the overabundance of GPS sources in the local Universe relative to the number of large galaxies. Additionally, the steep optically thin component of \pkssrc suggests it is possible that part of the USS population consists of a population of dying GPS sources for which the spectral break has shifted below the low radio frequency turnover.

\section*{Acknowledgments}

The authors thank Geoffrey Bicknell, David Jauncey and Edward King for stimulating discussions about the theoretical aspects of the absorption models and the VLBI measurements of PKS~B0008-421. This scientific work makes use of the Murchison Radioastronomy Observatory, operated by CSIRO. We acknowledge the Wajarri Yamatji people as the traditional owners of the Observatory site. Support for the MWA comes from the U.S. National Science Foundation (grants AST-0457585, PHY-0835713, CAREER-0847753, and AST-0908884), the Australian Research Council (LIEF grants LE0775621 and LE0882938), the U.S. Air Force Office of Scientific Research (grant FA9550-0510247), and the Centre for All-sky Astrophysics (an Australian Research Council Centre of Excellence funded by grant CE110001020). Support is also provided by the Smithsonian Astrophysical Observatory, the MIT School of Science, the Raman Research Institute, the Australian National University, and the Victoria University of Wellington (via grant MED-E1799 from the New Zealand Ministry of Economic Development and an IBM Shared University Research Grant). The Australian Federal government provides additional support via the CSIRO, National Collaborative Research Infrastructure Strategy, Education Investment Fund, and the Australia India Strategic Research Fund, and Astronomy Australia Limited, under contract to Curtin University. We acknowledge the iVEC Petabyte Data Store, the Initiative in Innovative Computing and the CUDA Center for Excellence sponsored by NVIDIA at Harvard University, and the International Centre for Radio Astronomy Research (ICRAR), a Joint Venture of Curtin University and The University of Western Australia, funded by the Western Australian State government. The Australia Telescope Compact Array is part of the Australia Telescope National Facility which is funded by the Commonwealth of Australia for operation as a National Facility managed by CSIRO. This paper includes archived data obtained through the Australia Telescope Online Archive (http://atoa.atnf.csiro.au). This research has made use of the NASA/IPAC Extragalactic Database (NED) which is operated by the Jet Propulsion Laboratory, California Institute of Technology, under contract with the National Aeronautics and Space Administration.

\bibliographystyle{apj} 
\bibliography{callinghampks0008-421.bbl}
	
\end{document}